\documentclass[prb,twocolumn,showpacs,floatfix]{revtex4}
\usepackage{graphicx}
\usepackage{dcolumn}
\usepackage{bm}

\begin{document}

\preprint{}

\title{Scattering on the lateral one-dimensional superlattice with spin-orbit coupling}

\author{D.V. Khomitsky}
 \email{khomitsky@phys.unn.ru}
 \affiliation{Department of Physics, University of Nizhny Novgorod,
              23 Gagarin Avenue, 603950 Nizhny Novgorod, Russian Federation}

\date{\today}

\begin{abstract}
The problem of scattering of the two-dimensional electron gas on the lateral one-dimensional superlattice
both having different strengths of Rashba spin-orbit coupling is investigated. The scattering is considered
for all the electron states on a given Fermi level. The distribution of spin density components along
the superlattice is studied for the transmitted states where the formation of standing waves is observed.
It is found that the shape of spin density distribution is robust against the variations of the Rashba
coupling constants and the Fermi level in the electron gas.
\end{abstract}


\maketitle

\section{Introduction}

In two-dimensional semiconductor heterostructures the spin-orbit (SO) interaction is usually
dominated by the Rashba coupling\cite{Rash60} coming from the structure inversion asymmetry of
confining potential and effective mass difference. The interest to these structures is related
to the possible effects in charge and spin transport which produce novel ideas on the spin control
in semiconductor structures and give rise to the applications of spintronics.\cite{Zutic}
The idea to control the spin orientation in the beam of particles by means of SO coupling has been
proposed in terms of spin optics.\cite{Kho04,Sh05} In particular, the scattering on the border
of two half-spaces each having a different value of SO coupling constants was studied.\cite{Kho04}
It was shown that the spin orientation in transmitted wave strongly depends on the chirality of
the incident one as well as on the angle of incidence and the angles of total reflection exist.
Later the same authors applied their results for the case of spin polarizing in a system consisted
of ballistic and diffusive regions.\cite{Sh05} One of the possible ways to control
the band and the spin structure is to apply the gated structures with externally tuned periodic electric
potential. In our recent paper we studied quantum states and the electron spin distribution in
a system combining the spin-splitting phenomena caused by the SO interaction and the external
periodic electric potential.\cite{jetpl} In the present paper me make an extensive use of these results
for investigation of the problem of scattering for 2DEG with Rashba
SO coupling on the SO superlattice. We solve the scattering problem on the SO superlattice
occupying a half-space and study the transmitted states as a function of the Fermi energy of
the incoming states. For the transmitted states the space distribution of spin density components
is calculated for different values of Rashba coupling on both sides of the interface, for various
amplitudes of the Fermi level position in the 2DEG.

The paper is organized as follows. In Sec.II we formulate the scattering problem and describe
the incoming, reflected, and transmitted states. We also briefly discuss the structure of
the eigenstates of the SO superlattice. In Sec.III the space distribution of spin density in
the transmitted state is calculated, and different cases of Rashba coupling on both sides of
the interface are discussed. The concluding remarks are given in Sec.IV.

\section{The scattering problem}

\begin{figure}[t]
  \centering
  \includegraphics[width=85mm]{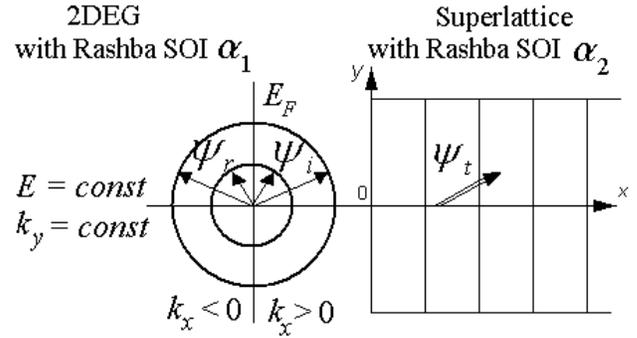}
  \caption{Geometry of scattering of 2DEG with Rashba SO interaction on the spin-orbit lateral superlattice.
           The incoming $\psi_i$ and reflected $\psi_r$ spinors are the eigenstates of Rashba
           Hamiltonian with spin-orbit coupling constant $\alpha_1$ and wavevectors belonging to
           the same Fermi contour. The transmitted states $\psi_t$ are the Bloch spinors
           corresponding to another spin-orbit coupling constant $\alpha_2$.}
  \label{scat}
\end{figure}

We consider the scattering of electrons with spin-orbit coupling constant $\alpha_1$ on the one-dimensional
superlattice occupying a half-space $x>0$ and also having a spin-orbit Rashba term with another value of
Rashba coupling constant $\alpha_2$. The incoming and reflected spinors are the eigenstates of Rashba
Hamiltonian and belong to the same Fermi energy of the 2D electron gas. The transmitted states are the Bloch
spinors with each of the components possessing the Bloch theorem. In addition to the energy, the $k_y$
component of the momentum is conserved since the system is homogeneous in the $y$ direction, as it is shown
schematically in Fig.\ref{scat}.

The half-space $x<0$ is the semiconductor structure with 2DEG characterized by
the effective mass $m$ and Rashba spin orbit coupling strength $\alpha_1$. The quantum
states here are the eigenstates of the Rashba Hamiltonian
$\hat{H}_0=\hat{p}^2/2m+ \alpha_1 (\hat{\sigma}_x\hat{p}_y-\hat{\sigma}_y\hat{p}_x)$
where $\hbar=1$. The eigenstates of this Hamiltonian are two-component spinors
$\psi_i=\psi_{{\bf k}\lambda}=e^{i{\bf kr}}\left(1, \quad  e^{i\theta}\right)/\sqrt{2}$
where $\lambda= \pm 1$ and $\theta={\rm arg}[k_y - ik_x]$. The energy of the state is
$E_0({{\bf k},\lambda})=\frac{k^2}{2m}+ \lambda \alpha_1 k$. It should be stressed that
this wavefunction does not exhibit any spin texture $S_i=\psi^{\dagger} {\hat \sigma}_i\psi$, i.e.
it determines a uniform space distribution of all spin density components $S_x=\lambda \cos \theta_0$,
$S_y=\lambda \sin \theta_0$, and $S_z \equiv 0$. The idea of the system setup in Fig.\ref{scat} is
to convert this uniform distribution into a non-trivial spin texture by using a superlattice.

The incoming state is scattered on the border of the SO superlattice occupying the area at $x>0$.
In the left part of the space $x<0$ there is the reflected state which is the linear combination of all
eigenstates of Rashba Hamiltonian with the same energy as the incoming state and with $k_x<0$.
The wavevector modules are equal to $k_{1,2}=\sqrt{2mE+(m\alpha)^2} \pm m\alpha_1$, and the $k_x$ component
for each $k_{1,2}$ at fixed $k_y$ is given by the usual relation $k_{1,2x}=\sqrt{k_{1,2}^2-k_y^2}$.
Thus, the reflected state at $x<0$ has the following form:

\begin{eqnarray}
x<0: \quad
\psi_{\rm r}=
r_1 \frac{e^{-ik_{1}x+ik_{y}y}}{\sqrt{2}}
\left(
\begin{array}{c}
1 \\
-e^{i\theta_1}
\end{array}
\right)
+
\nonumber
\\
+
r_2 \frac{e^{-ik_{2}x+ik_{y}y}}{\sqrt{2}}
\left(
\begin{array}{c}
1 \\
e^{i\theta_2}
\end{array}
\right).
\label{psiref}
\end{eqnarray}

\noindent Here the phases are defined by the momentum components as
$\theta_{1,2}={\rm arg}[k_y - ik_{1,2x}]$ and $r_{1,2}$ are the reflection coefficients
which will be found below.

On the right-hand side in Fig.\ref{scat} at $x>0$ the transmitted electrons travel trough
the SO superlattice. The transmitted state is the linear combination of the eigenstates of
the SO superlattice with the energy and $k_y$ equal to those of the incoming state:

\begin{equation}
x>0: \quad
\psi_{\rm t}=\sum_{j} c_j \psi(k_j,k_y)
\label{psit}
\end{equation}

\noindent where the coefficients $c_j$ can be found from the boundary conditions. The wavefunctions
$\psi(k_j,k_y)$ are the Bloch eigenstates of the Hamiltonian in the SO superlattice having
the form \cite{jetpl}

\begin{equation}
\psi_{s\bf k}=\sum_{\lambda n} a^s_{\lambda n} ({\bf k})
\frac{e^{i{\bf k}_n{\bf r}}}{\sqrt{2}} \left(
\begin{array}{c}
1 \\
\lambda e^{i\theta_n}
\end{array}
\right),
\quad \lambda = \pm 1
\label{wf}
\end{equation}

\noindent where $k_x$ is the quasimomentum in the 1D Brillouin zone $-\pi/a \le k_x \le \pi/a$,
$s$ is the band number, and $\theta_n={\rm arg}[k_y - ik_{nx}]$. The coefficients $a^s_{\lambda n}$
are found by diagonalization of the superlattice Hamiltonian in the basis of Rashba spinors.
The 1D superlattice potential in our problem can be chosen in the simplest form $V(x)=V_0 \cos (2\pi x /a)$
where $a$ is the superlattice period and $V_0$ is the potential strength.

The scattering on the interface at $x=0$ is described by the boundary conditions.
For the problem considered in the paper these conditions have the form of the continuity equations
which follow from the Schr\"odinger equation and can be written as

\begin{eqnarray}
\psi \mid_{x=0-} = \psi \mid_{x=0+},
\\
\hat{v}_x\psi \mid_{x=0-} = \hat{v}_x\psi \mid_{x=0+}
\label{cont}
\end{eqnarray}

\noindent where the velocity operator

\begin{equation}
\hat{v}_x=\frac{\partial \hat{H}}{\partial k_x} =\frac{\hat{p}_x}{m}-\alpha {\hat \sigma}_y.
\label{vx}
\end{equation}

\noindent The equations (\ref{cont}) link the wavefunction $\psi_{\rm i}+\psi_{\rm r}$ at the left
half-space $x<0$ and the wavefunction $\psi_{\rm t}$ at the right half-space $x>0$.
Since both of the equations in (\ref{cont}) are written for two-component spinors, one
has a system of four algebraic inhomogeneous equations describing the scattering which
can be easily solved.

The quantum numbers which remain to be good during the scattering on 1D superlattice
are the $k_y$ component of the momentum and the energy of the incoming state.
Here one has to distinguish the case when the energy of the incoming state at fixed
$k_y$ is within the limits of one of the superlattice bands and when this energy
corresponds to a gap in the superlattice spectrum. The first case corresponds
to the solution of system (\ref{cont}). For the second case the solution to the Schr\"odinger equation
is not finite on the whole $x$ axis and thus there are no states which propagate from the scattering
interface through the superlattice. We call such case as a case of total reflection in analogy with optical
scattering. It should be mentioned that such effect was already observed for the scattering of the Rashba
states on the interface between two areas with different SO constant.\cite{Kho04}
The states which do not propagate through the superlattice and are localized at the
interface border are known as Tamm states. Such states were studied previously both
in bulk crystals \cite{Goodwin,Kevan} and later in the superlattices.\cite{Ohno,Sy,Klos}
In the latter case it was shown that typically the Tamm states decay inside the superlattice
on the length of several periods with different results varying from two - three \cite{Ohno}
to five - seven \cite{Klos} lattice periods. In our case these results mean that
the typical penetration length of Tamm states will be of the order of 100 - 700 nm
which is substantially smaller than the total length of superlattices actually used in
the present experiments. Hence, there will be no detection of such states with the possible device
mounted after the superlattice. Thus, we neglect the Tamm states localized at the interface
and consider only the Bloch states with the energy belonging to the bands of the
superlattice which were discussed above.

\section{Spin texture of the transmitted state}

When the transmitted state (\ref{psit}) is fully determined, one can calculate the space distribution
of the spin density $\psi^{\dagger} {\hat \sigma}_i \psi$ for the transmitted state which depends on
the wavevector and polarization of the incident state. In a real experimental setup of
2DEG structure the electrons occupy not a single state with a given wavevector and polarization but
all of the states on the Fermi level, as it is shown schematically in Fig.\ref{scat}.
The electrons with $k_x>0$ travel to the scattering interface and take part in the scattering process.
Thus, it is reasonable to calculate the spin density for all the electrons with
a chosen Fermi energy and $k_x>0$ giving us the spin density distribution which can be
actually probed by a detector,

\begin{equation}
S_{i}(x,y)=\int_{k_{Fx}>0}\psi_{\rm t}^{\dagger} {\hat \sigma}_i \psi_{\rm t}dk.
\label{spins}
\end{equation}

\noindent Since the system is homogeneous in the $y$ direction, one may consider only the $x$
-dependence of (\ref{spins}) which may show some non-trivial spin texture along the superlattice.
As it was mentioned above, the incident state of 2DEG with Rashba spin-orbit coupling has
a space-independent spin density distribution. Below we shall see that a non-uniform
spin density distribution which can be actually probed by a detector may be created by scattering
on the SO superlattice.

\begin{figure}[ht]
  \centering
  \includegraphics[width=80mm]{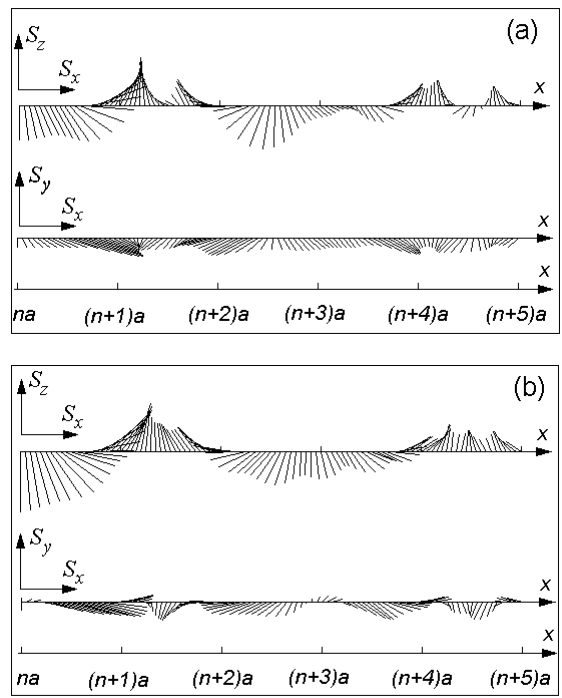}
  \caption{Spin texture along the superlattice for the Rashba constant $\alpha_2=3 \cdot 10^{-11}$ eVm
  inside and $\alpha_1=0.1 \alpha_2$ outside the superlattice. The periodic potential amplitude $V_0=5$ meV
  and the Fermi energy is (a) $E_F=10$ meV and (b) $E_F=30$ meV.}
  \label{sls}
\end{figure}

First, let us consider a case when the Rashba coupling constant $\alpha_1$ in the 2DEG on the left
is substantially smaller than the parameter $\alpha_2$ in the superlattice. This situation corresponds,
for example, to the GaAs-based structure attached to the InAs-based SO superlattice.
The results for the spin density distribution along the superlattice for $\alpha_1=0.1 \alpha_2$
are shown in Fig.\ref{sls} for the amplitude of the periodic potential $V_0=5$ meV and
for the values of the Fermi energy $E_F=10$ meV and $E_F=30$ meV of the incident state.
The upper plot on each figure shows the $(S_x,S_z)$ projections of the spin density (\ref{spins}) while
the lower one demonstrates the space dependence of $(S_x,S_y)$ components. The space distance on the plot
is measured in units of superlattice period $a=60$ nm and starts at $n\gg 1$ which means that
the spin detector is located far away from the superlattice border.
The spin texture in Fig.\ref{sls} has several remarkable features. First of all,
it has a non-zero component $S_z$ which is absent in spin density of the uniform 2DEG
with Rashba SO coupling. As for the spin expectation values $\sigma_i=\int S_i dx$ for
our problem, one has in general $\sigma_x=\sigma_z=0$ and $\sigma_y \ne 0$ which follows from the symmetry
considerations of the system (see Fig.\ref{scat}). Indeed, the system is symmetrical with respect to $y$
sign reversal which means for Rashba SO coupling that $\sigma_x=0$. The Rashba SO interaction also can not
create the $z$ polarization of 2DEG and thus $\sigma_z=0$, as in the initial state. It should be noted that
a similar feature was observed previously for the eigenstates in the SO superlattices at given quantum
numbers $(k_x,k_y)$ in the Brillouin zone.\cite{jetpl} The only symmetry breaking caused by the scattering
interface cancels the $x$ sign reversal symmetry, making only the states with $k_x>0$ to be actually
scattered. Thus, one can see in Fig.\ref{sls} and below in Fig.\ref{slb} that one sign of $S_y(x)$
dominates, leading in those cases to a nonzero expectation value $\sigma_y$. The other reason is
that the contributions to the spin expectation value $\sigma_y$ from two parts of Fermi contours
of the Rashba bands with $\lambda=\pm 1$ (see Fig.\ref{scat}) do not compensate each other due to
the distance $2m\alpha_1$ between the Fermi radii.
Another interesting feature of the spin density distribution in Fig.\ref{sls} is that it does not repeat
itself on the distance of one superlattice period. The explanation is that the transmitted state
(\ref{psit}) consists of the Bloch spinors with different $k_x$ components of the quasimomentum providing
the different partial wavelengths. As one can see from Fig.\ref{sls}, the approximate space period for
the spin density is about several superlattice periods and, as our calculations have shown, does not depend
on particular starting point $x=na$ if the condition $n\gg 1$ is satisfied. The latter means that the spin
density detector is located far away from the scattering border, as it is supposed to be in real experiments.
This circumstance allows to neglect the influence of the second right-hand border of the superlattice
while solving the scattering problem.

\begin{figure}[ht]
  \centering
  \includegraphics[width=80mm]{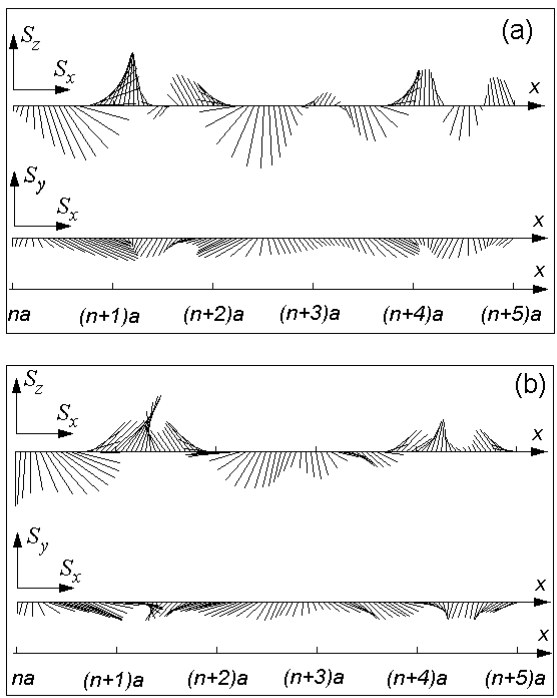}
  \caption{Spin texture along the superlattice for the Rashba constant $\alpha_1=3 \cdot 10^{-11}$ eVm
  outside and $\alpha_2=0.1 \alpha_2$ inside the superlattice. The periodic potential amplitude $V_0=5$ meV
  and the Fermi energy is (a) $E_F=10$ meV and (b) $E_F=30$ meV.}
  \label{slb}
\end{figure}

Now we turn our attention to the opposite case $\alpha_2=0.1 \alpha_1$ which can be realized experimentally,
for example, by the GaAs-based SO superlattice attached to the InAs-based 2DEG. The results for
the spin density distributions are presented in Fig.\ref{slb}.
Again one can see the similarity between all the spin density textures in Fig.\ref{slb} and
Fig.\ref{sls}. The integral spin density distribution (\ref{spins}) maintains qualitatively the same form
for different values of system parameters since it is sensible only to the global
characteristics of the energy spectrum of the superlattice which remain unchanged under variation of
the Fermi level position and Rashba coupling strength. We have also observed that the results presented above
are qualitatively the same for different values of the superlattice potential. Such robust spin density
shape indicates that the effects discussed in the paper should survive under various
perturbations which were left out of the scope in the present work such as defects and finite
temperature. This conclusion can be justified further if we mention that the energy scale
of the problem studied above belongs to the interval of $10\ldots 30$ meV,
which means that the effects discussed in the paper should be clearly observable at helium, and
possibly also at nitrogen temperatures.

\section{Conclusions}

We have studied the scattering of two-dimensional electron gas on the one-dimensional superlattice
where the spin-orbit coupling was taken into account for both systems. The space distribution of spin
density components was calculated for different values of Rashba coupling on both sides of the interface
and for various Fermi level position. The observed shape of spin density standing waves is found to be
insensitive to particular values of the electron Fermi energy and Rashba coupling strength indicating that
the effects discussed in the paper should survive under various perturbations such as defects and finite
temperature. The scale of energy involved in the processes discussed in the paper makes the results
to be promising for experimental observation.

\section*{Acknowledgments}

The author thanks V.Ya. Demikhovskii for numerous fruitful discussions.
The work was supported by the RNP Program of the Ministry of Education and Science RF,
by the RFBR, CRDF, and by the Foundation "Dynasty" - ICFPM.

\end{document}